# Application of relative configurational entropy as a measure of spatial inhomogeneity to Co/C film evolving along the temperature

Z. GARNCAREK
*Institute of Mathematics, University of Opole, Oleska 48, PL-45052 Opole, Poland*

T. MAJCHERCZYK
*Institute of Chemistry, University of Opole, Oleska 48, PL-45052 Opole, Poland*

D. POTOCZNA-PETRU
*Institute of Low Temperature and Structure Research Polish Academy of Sciences, Okolna 2, PL-50950 Wroclaw, Poland*

R. PIASECKI
*Institute of Chemistry, University of Opole, Oleska 48, PL-45052 Opole, Poland*
*E-mail: piaser@uni.opole.pl*

It is known that physical properties of a thin metallic film depend on spatial distribution of metal phase. Thermally treated thin film with metal particle clusters evolving along the temperature is a system complex enough to attract scientific interest, see for example [1]. However, the simple visual study of a transformation process of heated thin metallic films on the basis of the transmission electron microscopy (TEM) micrographs cannot give valuable information. For instance, the important (in context of possible connection with effective properties of the whole structure) feature of a given microstructure is its spatial inhomogeneity depending by definition on the length scale considered. Thus, we need a tool for quantitative characterization of the spatial inhomogeneity. A simple mathematical approach [2] related to a point object distribution has been already used to analyze an inhomogeneity degree of TEMs of thin gold films [3] on the basis of Doremus' micrographs [4] and polymer/carbon composites [5]. However, for digitized images with a pixel as natural unit of length this approach works only for the length scales commensurate with a side length of the image. Recently, a useful physical measure based on configurational entropy has been proposed [6] that overcomes this difficulty. The extension of this entropic measure for ''finite-sized'' objects [7] compared with the ''normalized information entropy'' [8] shows even more details for structures statistically self-similar.

Among other entropic measures worked out to characterize random microstructures are the ''local porosity entropy'' [9] and the ''configuration entropy'' [10]. They are based on the adaptation of Shannon information entropy and were found to be rigorously connected [11].

In this letter we apply the simple in usage entropic measure for point objects [6]. A linear transformation $f(S)$ of configurational entropy $S$ with length scale dependent coefficients as a measure of spatial inhomogeneity has already been tested for computer generated pixel distributions. To check its behaviour on real micrographs the Co/C film evolving along the temperature was chosen. The method used can be treated as one possible approach to finding a supposed connection between the changes in the structure of a thin film subject to heating conditions and its effective properties depending on temperature.

To present the main idea of the method let us consider an electron micrograph of a thin metallic film. Its digitized binary image of size $L \times L$ can be treated as a set of $\chi = (L/k)^2$ lattice cells of size $k \times k$ in which $n$ black pixels considered as point objects are distributed. These pixels represent small metal grains on a photographic negative. The pixel clusters certainly reflect less or more accurately (in dependence of the scanner resolution used) the two-dimensional area covered by a metallic phase. For each length scale $k$ with a given object distribution $(n_1, ..., n_i, ..., n_\chi)$ one can associate a configurational entropy $S = k_B \ln \Omega$, where the Boltzmann constant will be set to $k_B = 1$ for convenience and $\Omega$ is the number of different ways of generating the fixed distribution of objects. To evaluate the spatial inhomogeneity as the average deviation from the most uniform object arrangement, defined by condition $| n_i - n_j | \leq 1$ for each pair $i \neq j$, we use $f(S) \equiv (S_{\max} - S)/\chi$. The highest possible value of configurational entropy $S_{\max}$ (at a given length scale) corresponds the most spatially uniform object configuration while $S$ relates to the actual configuration. A more detailed description is given in the earlier paper [6]. Here the final formula is presented

$$f(S) = -\frac{r_0}{\chi} \ln(n_0 + 1) + \frac{1}{\chi} \sum_{i=1}^{\chi} \ln\left(\frac{n_i!}{n_0!}\right),$$ (1)

where $r_0 = n \,[\text{mod } \chi]$ and $n_0 = (n - r_0)/\chi$ are used to find $\Omega_{\max}$ by fixing $n_0 + 1$ and $n_0$ objects in $r_0$ and $\chi - r_0$ cells, respectively. Interestingly, if $L \,[\text{mod } k] = 0$ then $f(S(k, L)) = f(S(k, mL))$, where the final pattern of size $mL \times mL$ is formed by $m^2$-fold periodical repetition of an initial arrangement. This simple property allows us to calculate the exact value of (1) also for the scales incommensurate with the side length $L$ by defining $f(S(k,L)) = f(S(k, m'L))$ with $(m'L)$ [mod $k$] = 0. The lowest $f(S)$ value equals to 0 and is always reached for $k = 1$ and $L$. Otherwise, $f(S(k)) \approx 0$ characterizes an object arrangement close to the most spatially uniform configuration at a given scale $k$. The highest possible value for a given length scale is attained when all objects are placed in one lattice cell. Such a case corresponds to the strongest deviation per cell from a possible maximally uniform distribution for the length scale considered. When a dominating shape of $f(S)$ appears within a range of length scales then the significant perturbations of the spatial homogeneity are present. They are caused mainly by the clustering processes of objects.

Thin cobalt films (2 nm thick) were fabricated by evaporation of pure cobalt under high-vacuum conditions ($10^{-6}$ Pa) onto the platinum microscope grids covered with carbon substrate. Thin film carbon substrates were prepared by vacuum deposition from carbon arc. Electron diffraction analysis of grids with carbon substrate showed only the presence of diffuse rings, indicating that the substrates were amorphous. Next, the Co/C films were heated in flowing purified $H_2$ at 673, 773 and 873 K for $4h$, respectively. After each stage of the thermal procedure, changes in the structure of the films were monitored by TEM using a Tesla BS 613 microscope. Three micrographs made at the Institute of Low Temperature Polish Academy of Sciences in Wroclaw were taken into account. All the micrographs were correspondingly put to the same enlargement (1 cm in the picture = 1000 Å).

Each of enlarged micrographs was transferred using a scanner into computer memory. In this manner the 256 grey scale images of $1453 \times 2600$ pixels were obtained. The resolution of 300 dpi has been used as optimal to reach detailed picture in binary versions as well as to save the computation time. The square of side length $L = 1080$ in pixels was chosen as appropriate working section with respect to the regularity of film coverage in four subsquares each of size $540 \times 540$. The final binary images with cobalt surface coverage $\varphi = 0.35$, 0.20 and 0.17 are showed in Figs 1−3.

The relative configurational entropy measure given by Eq. (1) was calculated for *each* length scale within considered $k$-interval [1, 200]. The results are summarized in Fig. 4. Considering the dominating parts of the curves (without the many local perturbations) is sufficient to reveal the most important information included in the ''shapes'' of the curves, see the inset in Fig. 4. For the investigated temperatures $T_A = 673$, $T_B = 773$ and $T_C = 873$ K we can distinguish the corresponding length scales $k_A \approx 46$, $k_B \approx 49$ and $k_C \approx 54$ in pixels, approximately equivalent to 38.5, 41 and 45.1 nm. In the vicinity of each of these scales the averaged perturbations of the spatial homogeneity in planar distribution of cobalt phase on carbon substrate are relatively the most significant. Moreover, the last two distinguished scales, $k_B$ and $k_C$, are clearly shifted in regard to the first one towards the larger lengths along the increasing heating temperature. This feature underlines the physical significance of the measure. Such a behaviour is a manifestation of cobalt aggregation (easily seen in Figs 2 and 3) caused by vacuum heating of the films. Similar aggregation is observed as we increase the deposition temperature. Then the film coverage decreases with the larger metal island sizes. A recent example of such behaviour is Pd grown on MgO(001) surface [12]. Another feature refers to the qualitative similarity of the discontinuous film structures in Figs 2 and 3 with clearly separated cobalt islands. Namely, the approximate relation can be extracted from the inset in Fig. 4 for the curves B and C in the vicinity of $k_B$ and $k_C$: $f_C(S(k + k_C - k_B)) \approx f_B(S(k)) + 0.3(T_C - T_B)$. Also, the more extended convexity around the length scale $k_A$ indicates for the greater dispersion in average size of cluster distribution for the maze structure in Fig. 1.

Concluding, for the system of our concern within a narrow range of length scales the measure $f(S)$ reveals a dominant increase of spatial inhomogeneity of cobalt phase with a shift along the sample heating temperature. It is worth emphasizing that our method can be also useful to study correlation between two components selected from temperature, metal and non-metallic substrate, when the value of third component is keeping constant. Despite the simplicity of the measure used it seems to be a sensitive tool for experimentalists to evaluate even the subtle differences of spatial inhomogeneity in thin metallic films.

**Figure Captions**

Fig. 1. Representative binary image of size 1080x1080 in pixels of Co/C film after heating at the temperature $T_A = 673$ K with cobalt surface coverage $\varphi = 0.35$.

Fig. 2. Similarly as Figure 1, representative binary image for $T_B = 773$ K and $\varphi = 0.20$.

Fig. 3. Same as Figure 1, for $T_C = 873$ K and $\varphi = 0.17$.

Fig. 4. The relative configurational entropy measure $f(S)$ as a function of length scale $k$ expressed in pixels for Co/C thin film after heating at the temperatures 673 (A), 773 (B) and 873 K (C). After removing the local perturbations, the inset clearly shows the shift of the dominating parts in the curves B (filled circles) and C (open triangles) in regard to the curve A (open circles).

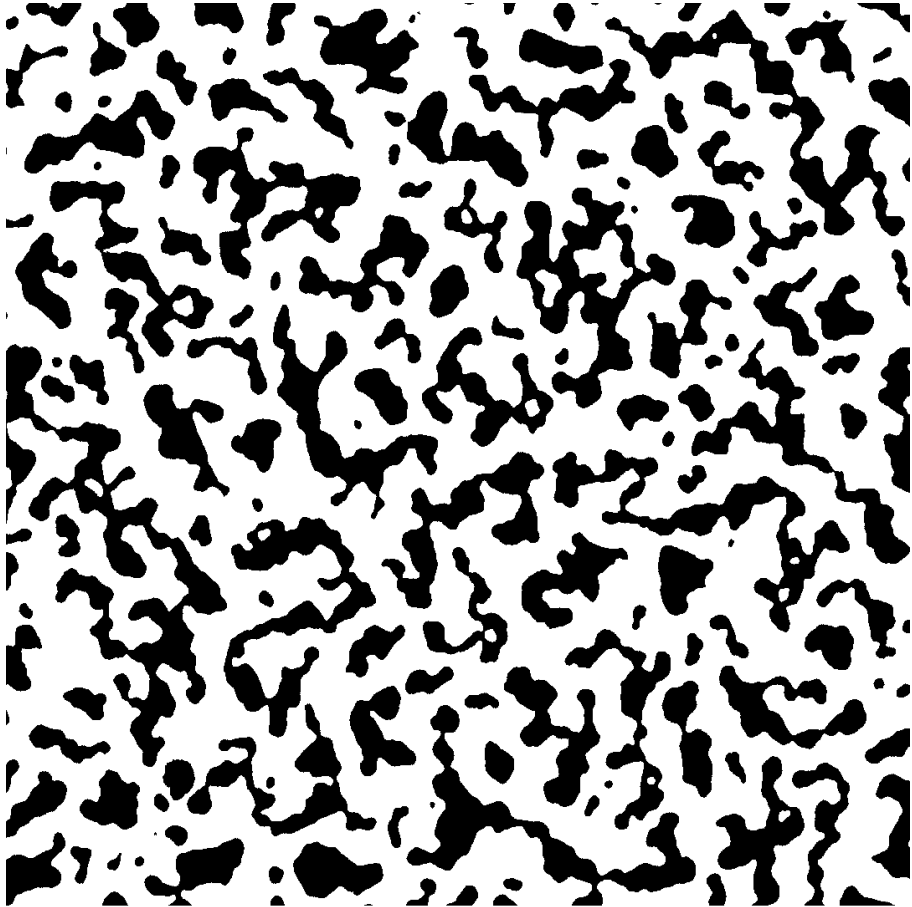

├──────┤ 100 nm

Fig. 1.

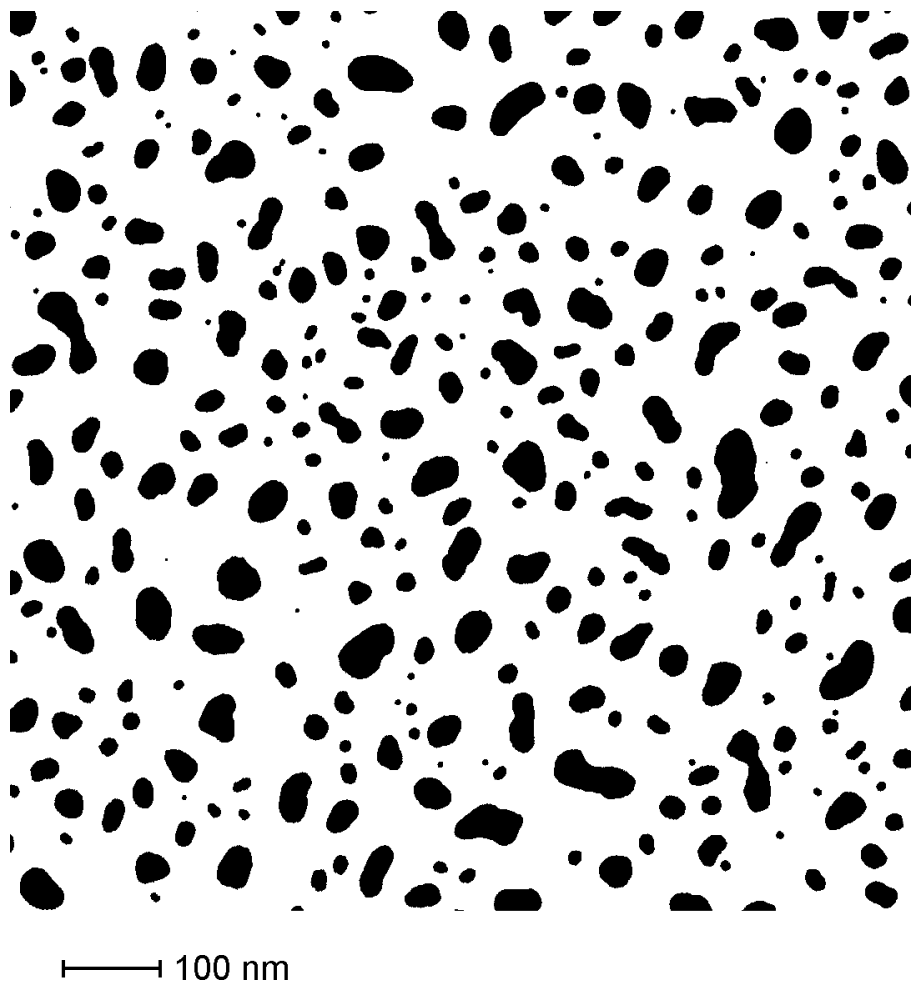

100 nm

Fig. 2.

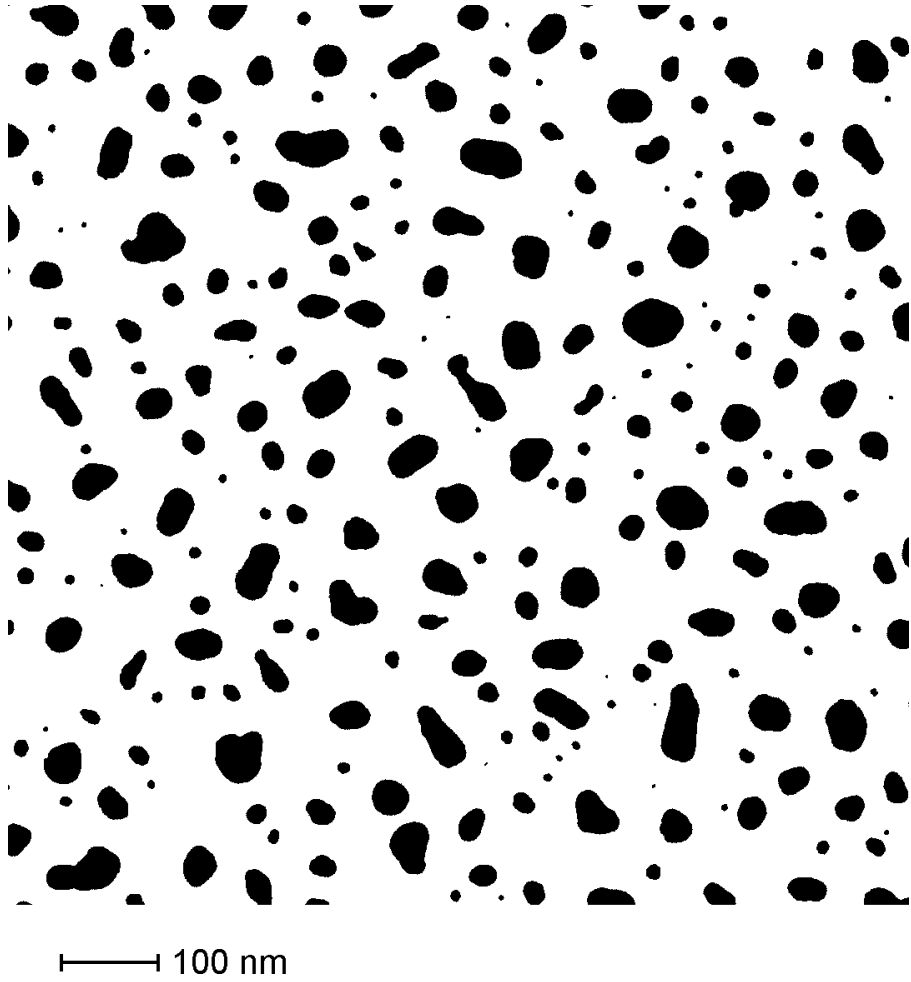

Fig. 3.

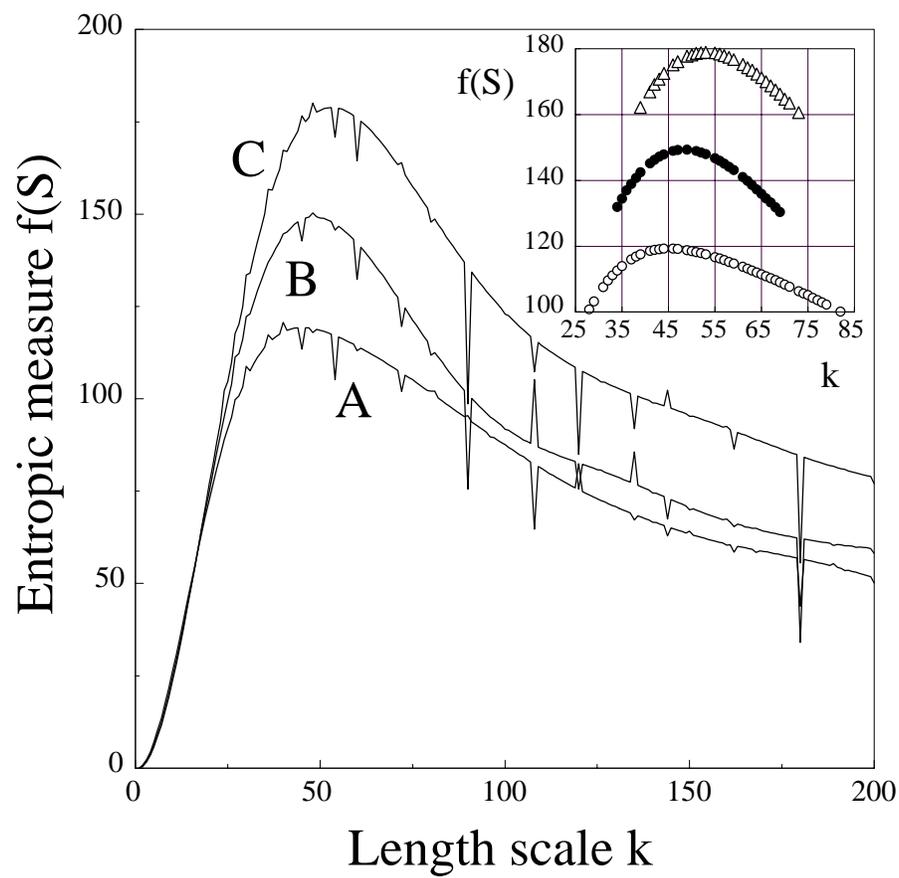

Fig. 4.